\documentclass[a4paper,twoside]{IEEEtran}
\usepackage{color}

\usepackage{amsmath,amssymb,graphicx,subfig}
\usepackage[natbib=true,
            giveninits=true,
            maxbibnames=9,
            sorting=none,
            style=numeric-comp,
            backend=bibtex,
            doi=false,
            isbn=false,
            url=false]{biblatex}
\renewbibmacro{in:}{}

\newcommand{\quotes}[1]{``#1''}

\usepackage[colorlinks=true,linkcolor=blue,urlcolor=black]{hyperref}
\usepackage{bookmark}
\begin{document}

\title{ Attenuation corrected activity reconstruction for listmode TOF-PET data }
\author{	
   Ahmadreza~Rezaei,
   Suleman~Surti, 
   Samuel~Matej, 
   Joel~S.~Karp, 
   Johan~Nuyts
   \thanks{ 
      This work was supported by the Research Foundation Flanders (FWO) G062220N and from NIH R01-CA113941.}
   \thanks{ 
      This work did not involve human subjects or animals in its research.}
   \thanks{
      A.~Rezaei (ahmadreza.rezaei@uz.kuleuven.be) and J.~Nuyts are with KU Leuven - University of Leuven, Department of Imaging and Pathology, Nuclear Medicine \& Molecular imaging; Medical Imaging Research Center (MIRC), B-3000, Leuven, Belgium.}
   \thanks{
      S.~Surti, S.~Matej and J~Karp are with Department of Radiology, Physics \& Instrumentation Group, University of Pennsylvania, Philadelphia, PA 19104-6112, USA. } 
\vspace{-1cm}
}
\maketitle

\begin{abstract}
Time-of-flight (TOF) positron emission tomography (PET) emission data allow joint estimation of the activity and attenuation images. 
The methods proposed so far rely on reconstructions from TOF histograms of the measurements. 
However, with the rapid data increase of total-body (TB) scanners and the ever improving TOF and intrinsic resolutions, more memory and computation efficient techniques for joint reconstruction need to be developed.
In this work, a modified version of the maximum likelihood activity and attenuation correction factors algorithm (MLACF) is presented, that jointly estimates the activity image together with the attenuation factors (AFs) from listmode emission data. 
The method does not require TOF histogramming of the data, however it still relies on a coarsely sampled (i.e. mashed) non-TOF histogram of the emission data.
The activity distribution is subsequently quantified by reconstructing and enforcing known expected tissue attenuation on an attenuation reconstruction from the estimated AF values.
The algorithm is evaluated using 3D simulations as well as clinical TOF-PET measurements of a patient scan and compared to reference reconstructions. 
As could be expected, we find the reconstruction accuracies improve with an \quotes{optimal} histogramming of the TOF emission data, increasing the signal to noise ratio of the data used for joint estimation.
With an \quotes{optimal} histogramming of the measurements, the attenuation corrected activity distributions from TOF emission data only are comparable to the gold-standard in PET/CT.
\end{abstract}


\section{Introduction}
The very high sensitivity of long axial field-of-view scanners \cite{karp2020j, badawi2019j} allows high quality tracer distribution reconstructions from short acquisitions, which as well as allowing investigations in multi-organ kinetics, is ideal for dynamic studies of fast kinetics. 
For studies of slower biologic processes however, the patients still need to be followed up over a longer time frame. 
It is for these studies that emission-based attenuation correction strategies are of paramount value for the patients to avoid the need for multiple whole-body CT scans. 
Additionally, another strong motivation for CT-less attenuation correction is for pediatric imaging, since there is a desire to eliminate the additional radiation burden on the children.

Since it was shown that the time of flight (TOF) PET data provide information about the attenuation that was not previously available \cite{defrise2012j}, the topic of joint activity and attenuation estimation has seen renewed interest and a number of newly developed algorithms have been proposed to exploit the added information available with TOF.
The majority of the developed methods aim at simultaneously reconstructing an activity and an attenuation image from the TOF-PET emission data \cite{rezaei2012j, Mehranian2015j, li2020j}. 
However, several methods have been proposed which avoid the reconstruction of the attenuation medium altogether and alternatively estimate the attenuation factors (AFs) which more directly influence the TOF emission measurements \cite{rezaei2014j, panin2012c}. 
Following the recent developments in convolutional neural networks a large collection of works have been proposed that study the network framework, its architecture and the inputs to obtain a map of the attenuation image which is ultimately used for attenuation/scatter correction in PET.
A nice review of these methods can be found in \cite{lee2021j}.

Typically, the emission data are stored as a sinogram (i.e. a histogram containing the number of events for each detector pair and each TOF bin). 
With increasing axial extents, improving TOF resolutions and the inclusion of depth of interaction information within the scintillating crystal, this memory-demanding  approach of storing the data is becoming less practical. 
A listmode file is an alternative way of storing the data, which becomes more efficient when the number of events is small compared to the total number of sinogram elements. 
Various event-based (listmode) reconstruction methods exist for standard activity reconstruction that avoid data histogramming (sinograms) and hence are better suited for the huge memory demands of total-body (TB) scanners. 
Joint reconstruction methods however, are not as well developed nor evaluated for reconstruction from listmode emission data.

We have previously proposed methods for joint reconstruction from listmode data \cite{rezaei2015c, rezaei2022c}. 
Although the method in \cite{rezaei2015c} was shown to be well suited to account for rigid patient motion during the scan, the method does not account for the additive contributions of randoms and/or scatter and computing the joint estimation scale factor from the estimated sensitivity image is non-trivial. 
In \cite{rezaei2022c}, we proposed an alternative, where the AFs were split into known water support and \quotes{residual} factor components, and where the AF estimation only required a non-TOF histogram of the emission data.
A Gaussian intensity prior was used on the estimated \quotes{residual} attenuation factors to resolve the joint reconstruction scale, and although this helped mitigate problems due to data-bins with limited information for AF estimation, it was later found to be insufficient to obtain quantitative reconstructions.
To mediate the computation demands of computing the sensitivity image with the recursively changing \quotes{residual} correction factors, the sensitivity image was computed once for all possible lines-of-response (LORs) using the water support correction factors and subsequently the emission data were pre-corrected for the estimated \quotes{residual} factor component.
Although pre-correcting the data for required corrections during activity reconstruction results in a relatively modest amplification of noise in standard reconstructions, we observed this to have a more severe effect on joint activity and attenuation reconstructions.
In this study, we provide a listmode formulation of the maximum likelihood activity and attenuation correction factor estimation (MLACF) technique which fully accounts for the additive contributions, avoids any pre-corrections to the emission measurements and computes the corresponding sensitivity image from the measured LORs as opposed to the list of all possible LORs.
Lastly, the estimated tracer distributions are made quantitative, by computing the appropriate joint reconstruction scale factor from a transmission reconstruction of estimated AFs, post MLACF reconstruction.

The paper is organized as follows: section \ref{methods} presents the listmode MLACF algorithm.
Section \ref{experiments} describes a Monte Carlo simulation experiment, where the joint activity and attenuation reconstructions can be compared to the corresponding ground-truth images. 
The section also describes the joint reconstruction from a real $^{18}$F-FDG whole-body patient scan, and the comparison of the jointly estimated activity image to a reference MLEM activity reconstruction.
The results of these experiments are shown and compared in section \ref{results}, and we conclude the paper by discussing the results in section \ref{discussion}.

\section{Methods\label{methods}}
We express the expected count $\bar y_{it}$ for line-of-response (LOR) $i$ and TOF-bin $t$ as:
\begin{eqnarray}
   \bar y_{it} = n_i a_i p_{it} + \bar s_{it} \label{model}
\end{eqnarray}
where $n_i$ is the detector-pair sensitivity, $a_i$ is the attenuation factor (AF), $\bar s_{it}$ is the expected additive contribution of randoms/scatter and 
\begin{eqnarray}
   p_{it} = \sum_j c_{itj} \lambda_j
\end{eqnarray}
is the TOF projection of the activity image $\lambda$, where $c_{itj}$ is the sensitivity of data bin $it$ to activity in voxel $j$.

In the following, we present the listmode MLACF activity update which is derived by maximizing the TOF log-likelihood for emission tomography accounting for the Poisson nature of the measurements.
On the other hand, for updating the attenuation factors, the Poisson distribution is approximated with a Gaussian distribution, such that the non-TOF log-likelihood becomes a simple non-TOF weighted quadratic difference \cite{panin2012c}.
This change eliminates the need for a TOF histogram of the measurements and provides a direct expression for the estimated attenuation factors as opposed to expression (3) in \cite{rezaei2014j}.

\subsection{Activity Update}
The TOF log-likelihood for emission tomography can be written as:
\begin{eqnarray}
L = \sum_{it} y_{it} \ln{\bar y_{it}} - \bar y_{it}. \label{lik}
\end{eqnarray}
Maximizing (\ref{lik}) with respect to the activity image and modifying the expression to accommodate listmode measurements, the listmode MLACF update is:
\begin{eqnarray}
\lambda_j^{m+1} = \frac{\lambda_j^m}{\sum_e c_{i_ej} n_{i_e} a_{i_e} / y_{i_e}} \sum_e c_{{(it)}_ej} n_{i_e} a_{i_e} / \bar y_{{(it)}_e} \label{act}
\end{eqnarray}
where $e$ defines the measured event, $i_e$ is the LOR associated with event $e$ and $y_{i_e} =  \sum_{ \{e'| i_{e'} = i_e\} } 1 $ equals the total number of events along the LOR.
We introduce here the convention that dropping the TOF-index $t$ denotes summation over the TOF bins, i.e $y_i = \sum_t y_{it}$.
Note that in MLACF $a_i=0$ if $y_i=0$ (see equation (3) in \cite{rezaei2014j}). 
As a result, the summation in the denominator is restricted to the LORs which have seen events.
In standard listmode MLEM reconstruction, the sensitivity image is computed from all possible LORs.

To efficiently obtain the non-TOF count $y_{i_e}$ from the TOF listmode data, a non-TOF mashed histogram (or sinogram) is created. 
The operation that is often referred to as mashing, i.e. combining multiple LORs into a single histogram bin, helps reducing the size of the required non-TOF histogram as well as prevents introducing bias in the activity reconstruction arising from the proposed way of computing the sensitivity image.

\subsection{Attenuation Update}
Assuming knowledge of the attenuating medium support, we split the AFs into two components $a_i = \alpha k_i r_i$ where $\alpha$ is the joint estimation scale, $k_i$ are the AFs corresponding to the known object filled with water attenuation and $r_i$ is the \quotes{residual} correction required for LORs that intersect media with a linear attenuation coefficient other than that of tissue e.g. lung/bone/bed/etc.
We absorb $k_i$ in $c_{itj}$, and use MLACF only to estimate $r_i$ (and $\lambda_j$).
Furthermore, to simplify the optimization, we approximate the logarithm of the Poisson likelihood with (data) weighted non-TOF squared differences. 
In addition, we add an intensity prior which encourages the $r_i$ to be close to unity, obtaining as the final cost function:
\begin{eqnarray}
Q = \sum_i \frac{1}{2} ( y_i - \bar y_i )^2 / y_i + \gamma R(r_i), \label{quad1}
\end{eqnarray}
where $R$ is an intensity regularizer and $\gamma$ is a hyper-parameter determining its strength.
The regularization of the estimated $r_i$ with the intensity prior drives the estimated values towards unity (no correction), limiting adverse effects otherwise observed in LORs with low or zero counts.

\subsubsection{$L_2$ Regularizer}
Using the quadratic penalty $R(r_i) = \gamma (r_i - 1)^2 / 2$ \cite{rezaei2022c}, the residual AFs are estimated for each event as:
\begin{eqnarray}
   r_{i_e} = \frac{(y_{i_e} - \bar s_{i_e}) + \gamma \, y_{i_e} / (n_{i_e} k_{i_e} p_{i_e})}{n_{i_e} k_{i_e} p_{i_e} + \gamma \, y_{i_e} / (n_{i_e} k_{i_e} p_{i_e})}. \label{att_l2}
\end{eqnarray}

\subsubsection{$L_1$ Regularizer}
Defining instead the penalty as $R(r_i) = \gamma |r_i - 1|$, the Iterative Shrinkage-Thresholding Algorithm \cite{beck2009j} is used to minimize this convex sub-problem.
The residual parameters are computed as:
\begin{eqnarray}
r_{i_e}^{m+1} \!\!\! &=& \!\!\! \mathbb{S}_{\gamma |.|} [ (r_{i_e}^m - 1) - \gamma \frac{\partial ( y_{i_e} - \bar y_{i_e} )^2 / y_{i_e}}{2 \partial r_{i_e}} ] + 1 \\
\!\!\! &=& \!\!\! \mathbb{S}_{\gamma |.|} [(r_{i_e}^m - 1) + (1 - \bar y_{i_e}/y_{i_e})n_{i_e} k_{i_e} p_{i_e} ] + 1 \label{att_l1}
\end{eqnarray}
where $\mathbb{S}_\epsilon [x]_i \! = \! max(|x_i|-\epsilon, \, 0) \, \, sign (x_i)$ is the soft-thresholding operator.

In both scenarios, the \quotes{residual} AF is estimated along the LOR of each recorded event, and can be treated as nuisance parameters in cases where quantification of the activity is not required. 
However, we often accumulate them into a sinogram, which can be useful when assessing the quality of the reconstruction. 
Furthermore, if quantification is desired, an attenuation image is reconstructed from the estimated attenuation sinogram to determine the global scale factor $\alpha$.

The estimation of correction factors along different LORs, is only possible given \quotes{sufficient} emission counts along the LORs.
Hence, data mashing for MLACF reconstructions of typical scan durations is essential.
Within our work, this mashing is applied where ever a histogram is required; i.e. for the non-TOF data $y_i$ in (\ref{act}), (\ref{att_l2}) and (\ref{att_l1}), and for $r_i$ in (\ref{att_l2}) and (\ref{att_l1}) in case quantification is required.

\subsection{Quantification}
The joint estimation scale $\alpha$ was computed (post MLACF) during an attenuation image reconstruction of the AFs using the maximum likelihood for transmission tomography (MLTR) \cite{vanslambrouck2014j}.
This was obtained by minimizing the quadratic differences of the estimated AFs from the reconstructed attenuation image obtained at each sub-iteration where the reconstruction was scaled to have expected tissue attenuation within a tissue region:
\begin{eqnarray}
R = \frac{1}{2} \sum_i (\alpha a^m_i - e^{-\sum_j l_{ij} \beta^m \mu^m_j} )^2  \label{quad2}
\end{eqnarray}
where $a^m_i = \alpha^m k_i r_i$, $\mu^m_j$ is the MLTR attenuation reconstruction at sub-iteration $m$ and $\beta^m$ is an intermediate scale enforcing expected tissue attenuation within the defined tissue mask.
Setting the derivative of (\ref{quad2}) to zero, we compute the joint estimation scale as:
\begin{eqnarray}
\alpha^{m+1} \!\!\!\!\! &=& \!\!\! \sum_i a^m_i e^{-\beta \sum_j l_{ij} \mu^m_j} / \sum_i (a^m_i)^2 \\
\!\!\!\!\! &\approx& \!\!\! \sum_i a^m_i e^{-\beta \sum_j l_{ij} \mu^m_j} / \sum_i e^{-2 \beta \sum_j l_{ij} \mu^m_j} \label{scale}
\end{eqnarray}
where we find approximation (\ref{scale}) to provide a more stable estimate of the scale with the noise present in $a^m_i$.
Iterating sufficiently long, $\alpha^m$ and $\beta^m$ converge to unity and the joint estimation scale is computed as $\alpha = \prod_m \alpha^m$.

\begin{figure}[t!]
	\centering
%
   \includegraphics[width=\columnwidth]{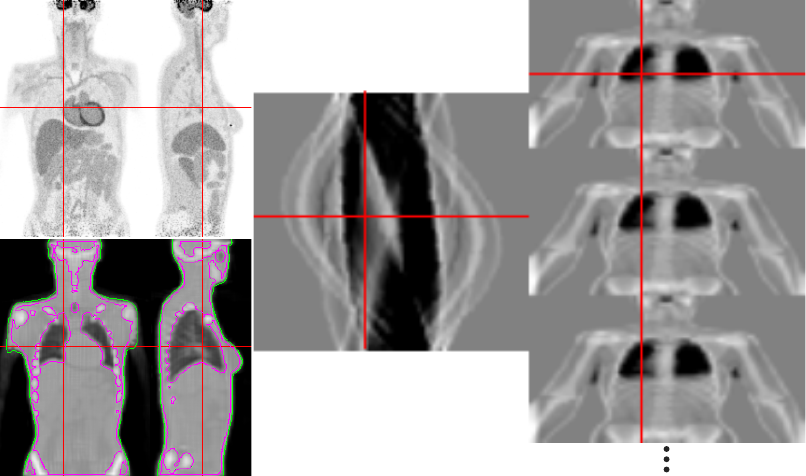}
	\caption{\label{xcat}
     Top-left: reference activity reconstruction (OSEM 3 iterations and 18 subsets) of the XCAT simulation data.
	  Bottom-left: reference attenuation reconstruction (OSTR 5 iterations and 18 subsets) of the mashed (mashing factor: 4) attenuation factors.
	  The green and cyan contours define the support and tissue region masks.
	  Right: a sinogram (and three projections with different axial obliqueness of the LORs) of the reference residual AF computed as the ratio of the AFs and the known water support AFs.}
\end{figure}

\begin{figure}[t!]
	\centering
%
   \includegraphics[width=\columnwidth]{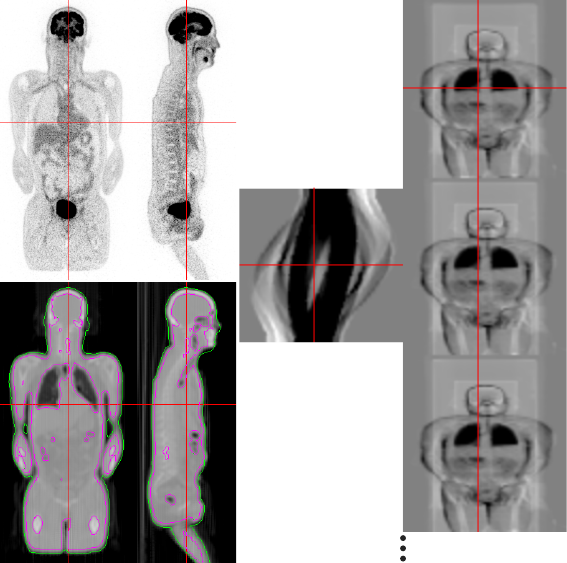}
	\caption{\label{patient}
     OSEM activity (top-left) and OSTR attenuation (bottom-left) reconstructions of the patient data, obtained after 3 and 5 iterations, respectively with 18 subsets.
	  A sinogram (and three projections with different axial obliqueness of the LORs) of the reference residual AF computed as the ratio of the CT AFs and the known water support AFs is shown on the right.
      }
\end{figure}

\begin{figure*}[ht]
	\centering
%
   \includegraphics[width=\textwidth]{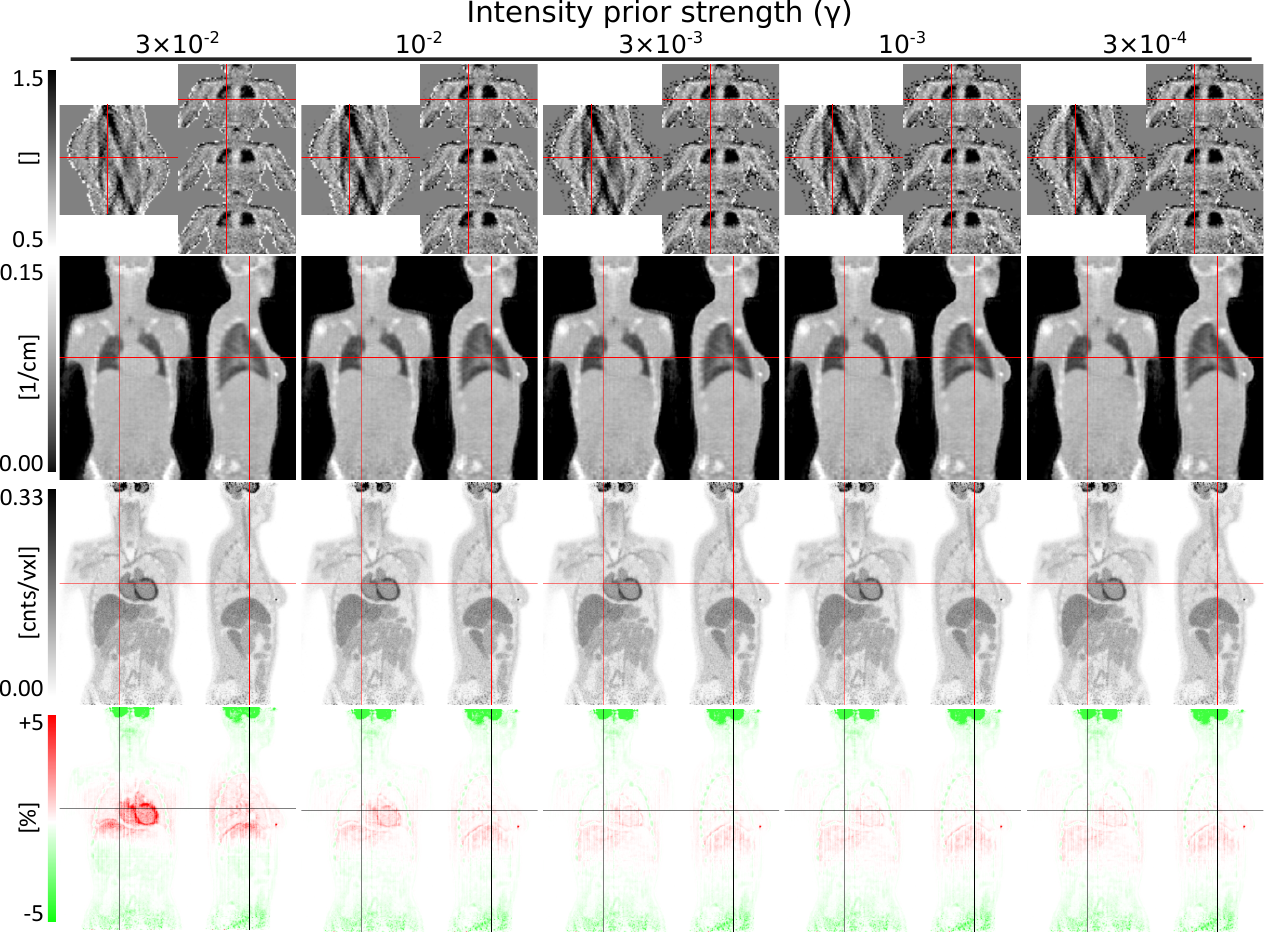}
	\caption{\label{xcat_trues}
      Simultaneous reconstruction results of the trues only experiment for simulated 3D XCAT phantom data.
      The estimated residual AFs (first row), MLTR attenuation (second row) and MLACF activity (third row) reconstructions as well as relative difference images (fourth row) to the ground-truth OSEM reconstruction figure \ref{xcat} of 3D XCAT phantom simulation with decreasing intensity prior strength on the estimated AFs (left-to-right).
      Difference images are shown at $\pm 5\%$ of the maximum activity color table.}
\end{figure*}

\begin{figure*}[ht]
	\centering
%
   \includegraphics[width=\textwidth]{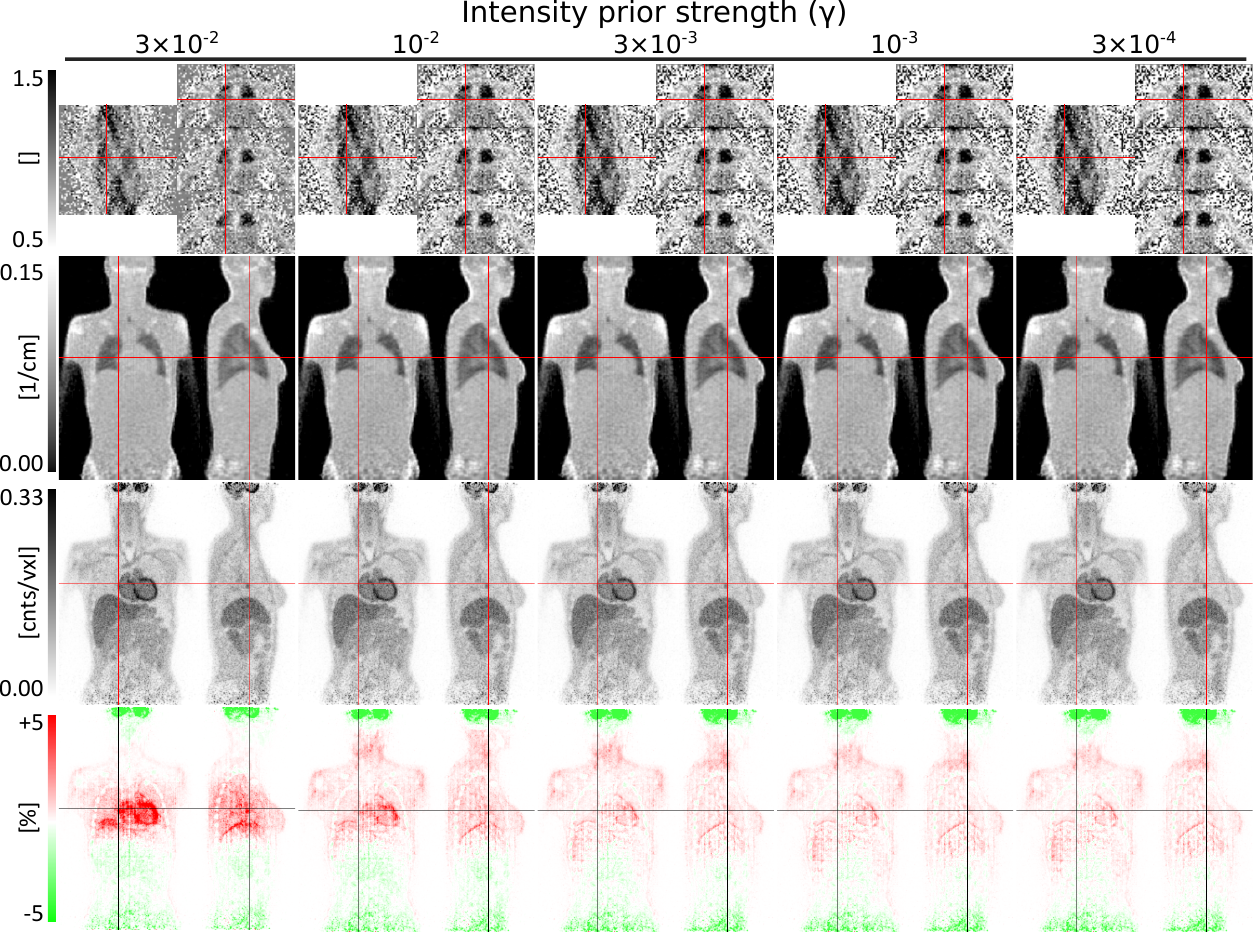}
	\caption{\label{xcat_recon}
      Simultaneous reconstruction results for simulated 3D XCAT phantom data with the inclusion of the additive randoms/scatter contribution.
      The figure is organized similar to figure \ref{xcat_trues}. }
\end{figure*}

\begin{figure*}[ht]
	\centering
%
   \includegraphics[width=\textwidth]{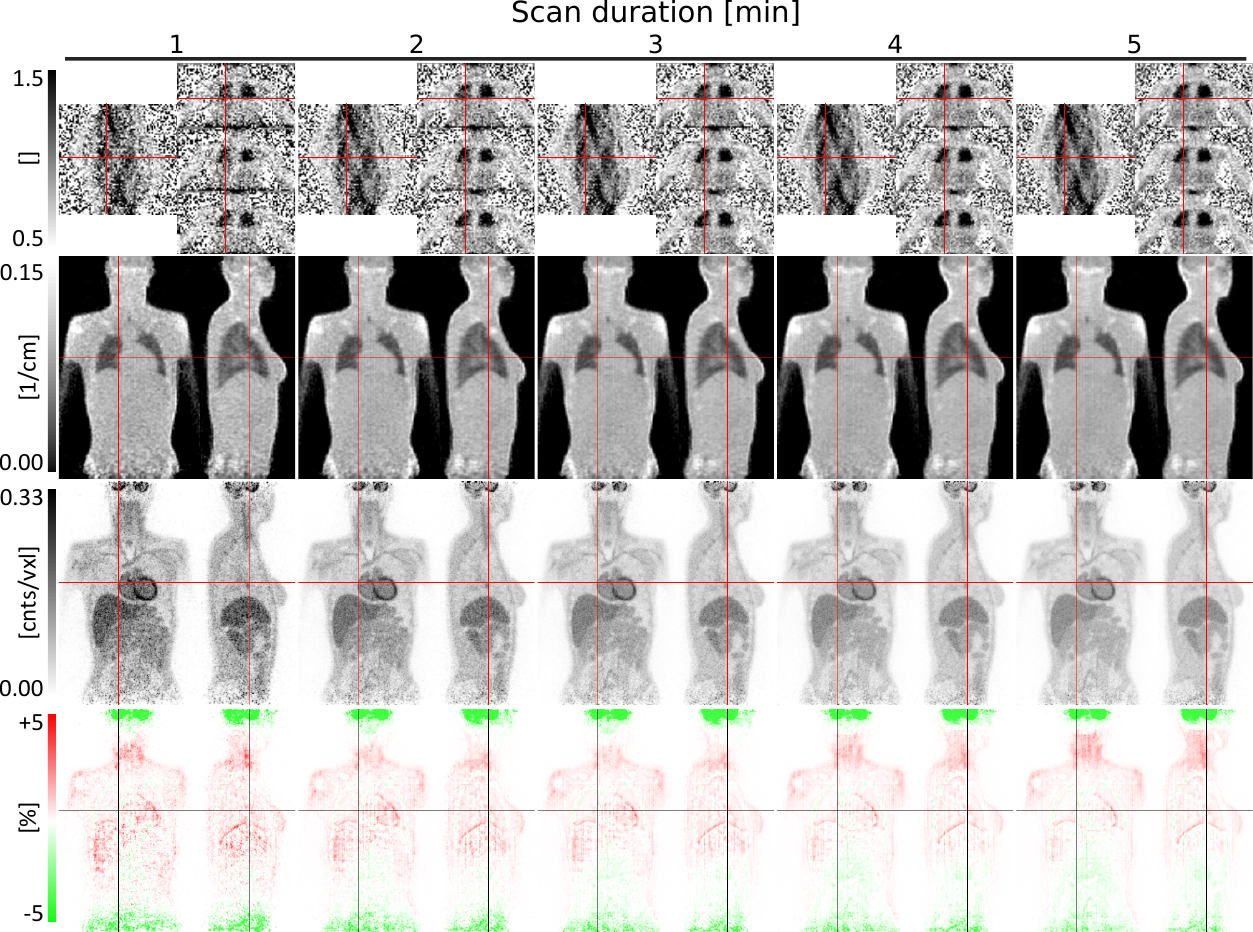}
	\caption{\label{xcat_noise}
      Simultaneous reconstruction results for simulated 3D XCAT phantom data for different levels of noise within the data ranging from 1 min to 5 min (left to right) of the simulated measurements. 
      The intensity prior strength was set to 3$\times$10$^{-3}$ in all reconstructions.
      The figure is organized similar to figure \ref{xcat_trues}. }
\end{figure*}

\begin{figure*}[ht]
	\centering
%
   \includegraphics[width=\textwidth]{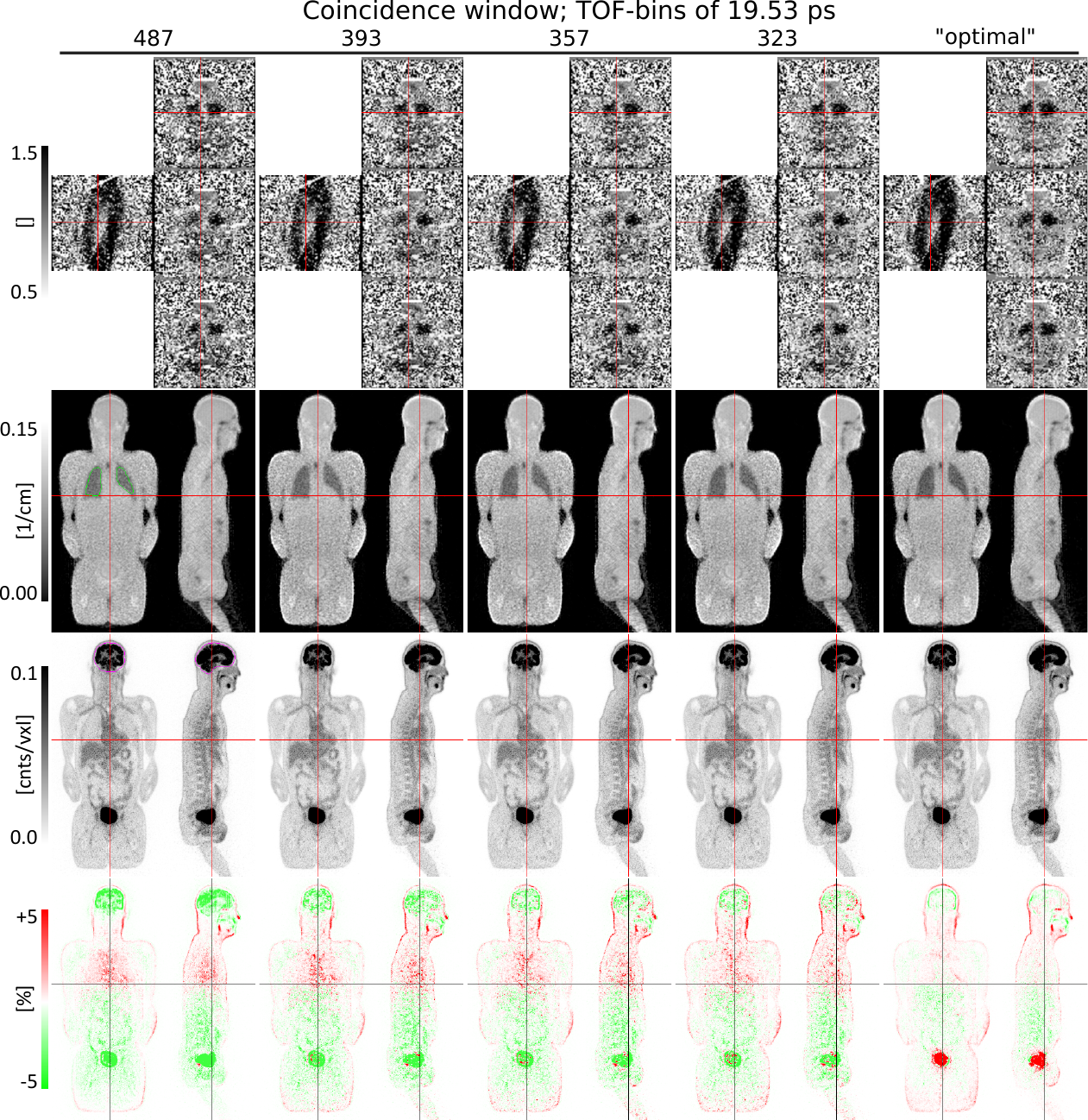}
	\caption{\label{patient_snr}
      Simultaneous reconstruction results for the patient data with a decreasing coincidence window and therefore with an increasing signal to noise of the data for attenuation estimation (left-to-right).
      The figure shows the estimated residual AFs (first row), MLTR attenuation (second row) and MLACF activity (third row) reconstructions as well as relative difference images (fourth row) to the gold-standard OSEM activity reconstruction using CT AFs. }
\end{figure*}

\begin{figure*}[ht]
	\centering
%
   \includegraphics[width=\textwidth]{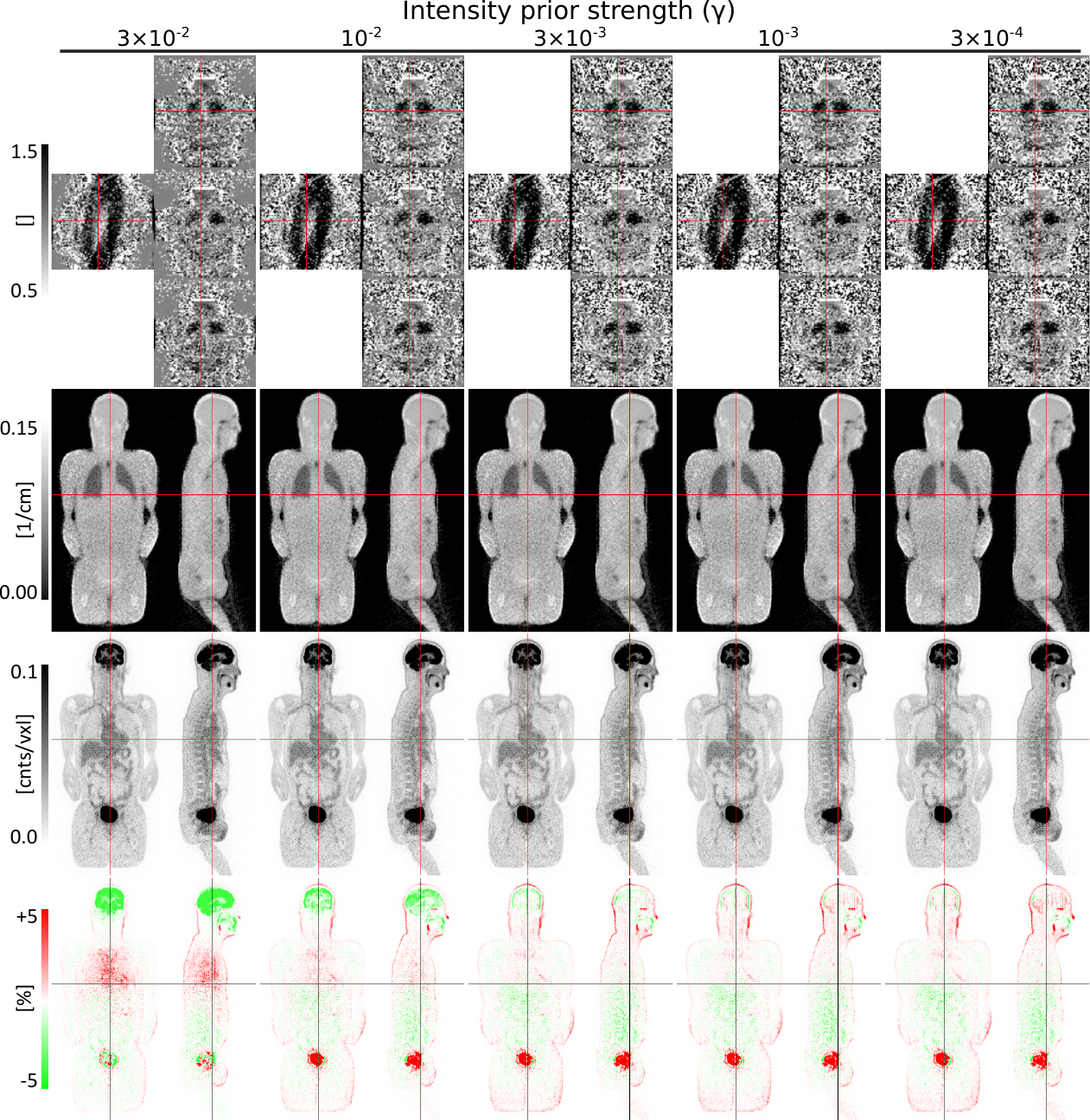}
	\caption{\label{patient_recons}
      Simultaneous reconstruction results for the patient data with decreasing intensity prior strength on the estimated AFs (left-to-right).
      The figure is organized similar to figure \ref{patient_snr}. }
\end{figure*}

\section{Experiment Design \label{experiments}}
The UPennPET Explorer \cite{pantel2019j} geometry and simulation parameters were used in the MonteCarlo simulation study, which consisted of 576 and 336 isotropic crystals of 4.1 mm in the transaxial and axial dimensions, respectively.
The scanner has a TOF resolution of 250 ps FWHM and the data are discretized in bins of 19.53 ps.
In both studies, non-TOF sinograms where generated with a crystal mashing factor of 4 (in transaxial and axial directions).
An expected randoms contribution was computed from the recorded delayed events using a maximum likelihood algorithm for delayed events by initially estimating a crystal map of the singles rates \cite{badawi1999j} (more information can be found in the Appendix), and a (sinogram) plane-dependent scaling of a simulated single scatter estimate \cite{rezaei2017j} was used as the scatter expectation.

The algorithm was initialized with a single iteration of the listmode OSEM algorithm using 18 subsets, and a \quotes{residual} AF sinogram of all ones. 
The reconstructions were repeated with different values for $\gamma$, the weight of the intensity prior, and the AF estimates (10 updates per sub-iteration) using the $L_1$ regularizer update (\ref{att_l1}) were restricted to the range of $[0.1,10]$ by truncating values outside this range.
MAPTR reconstructions of the AFs were subsequently made making use of the relative-difference prior \cite{nuyts2002j} (setting the weight and its edge tolerance to 10 and 1/10 of expected tissue coefficient, respectively) encouraging smoothness of the attenuation reconstructions.
A blank scan of ones and $\log{(10)}$ (chosen heuristically) within and outside the support was used to suppress the buildup of attenuation outside the activity support, akin to the missing attenuation data solution in \cite{nuyts1999j}.
Standard listmode OSEM activity reconstructions (3 iterations and 18 subsets) were then made with the scaled AF values in order to compare the reconstructions to the gold-standard reconstructions using CT AFs at similar convergence and noise levels.
Throughout the manuscript a white-to-black and black-to-white color map are chosen to show activity reconstructions and attenuation images/correction factors, respectively.

\subsection{Monte Carlo simulation}
Activity and attenuation images of the XCAT phantom were used for the MonteCarlo simulation study. 
Data were collected from 3 (of the 6) rings of the PennPET Explorer, and hence the activity was reconstructed in a 280 $\times$ 280 $\times$ 284 isotropic voxel grid of 2.5 mm.
A transaxial/axial crystal mashing factor of 4 was used to reduce the size of the required non-TOF sinogram to 77 $\times$ 72 $\times$ 1764 elements in the radial, angular and planar dimensions.
Listmode MLACF reconstructions were made from trues only events in addition to the prompts to analyze the reconstructions with and without the added uncertainty due to the additive randoms/scatter contributions.
Data from 2 min of the simulation were used in this study, where the non-TOF prompt sinogram had total and maximum counts of 276.0 $\cdot$ 10$^6$ and 166 events, respectively. 
Randoms and scatter fractions were computed as $n_r / n_y$ and $n_s / (n_y - n_r)$, respectively, where $n_r$ and $n_s$ are the total number of expected randoms and scattered events, and $n_y$ is the total number of recorded prompts and calculated to be 54.7$\%$ and 25.6$\%$, respectively.
The MLACF activity reconstructions are then compared to MLEM activity reconstructions with varying intensity prior strengths. 
Figure \ref{xcat} shows the ground-truth OSEM (3 iterations and 18 subsets) activity reconstructions as well as the residual AFs together with an OSTR (5 iterations and 18 subsets) of the attenuation estimate when the trans/axial mashing is applied to the AFs.

\subsection{Patient data}
Data were collected from all 6 rings of the PennPET Explorer, and the reconstruction volume was set to a 280 $\times$ 280 $\times$ 572 voxel grid with an isotropic voxel size of 2.5 mm. 
A non-TOF mashed (mashing factor 4) sinogram was generated which had 77 $\times$ 72 $\times$ 7056 elements in the radial, angular and planar dimensions, respectively.
Events from a 60~s duration of a 10 min (50-60 min post-injection) scan are used for listmode MLACF reconstructions and are compared to standard activity reconstructions of OSEM for a patient injected with 9.3 mCi of $^{18}$F-FDG. 
The emission data were collected within a 10 ns coincidence window, over a total of 511 TOF-bins, from the 6 rings of the PennPET Explorer.
To study effects of the out of the field-of-view random events (along the TOF dimension) on the LOR attenuation factors, the measurements were subsequently truncated to 427, 393, 357, 323 TOF-bins together with an \quotes{optimal} data truncation, i.e. a coincidence window per LOR, adjusted according to the support of the patient body.
A ratio of $(n_y - n_d) / n_y$, where $n_d$ is the total number of recorded delayed events, was computed as 28.2$\%$, 29.9$\%$, 32.0$\%$, 34.1$\%$ and 38.5$\%$ for the aforementioned histogrammings, respectively.
Joint reconstructions were then obtained with an intensity prior strength of 3$\times$10$^{-3}$, chosen heuristically based on the results of the simulation experiments.
Lastly, results of joint activity and attenuation reconstructions are presented varying the strength of the intensity prior using the \quotes{optimal} histogramming of the emission data.
The non-TOF sinogram had total and maximum counts of 716 $\cdot$ 10$^6$ and 394 events, respectively, with randoms and scatter fractions of 61.5 $\%$ and 15.2 $\%$, respectively.
Figure \ref{patient} shows the gold-standard OSEM (3 iterations and 18 subsets) activity and OSTR (5 iterations and 18 subsets) attenuation reconstruction using CT AFs.

\section{Results\label{results}}

\subsection{Monte Carlo simulation}
Figure \ref{xcat_trues} shows results of the trues-only simulated data with varying intensity prior strengths favoring a correction factor of 1.
The difference images reveal that comparatively accurate activity reconstructions can be obtained with moderate intensity prior strengths.
However, close to the edge of axial field of view the differences are larger which could be attributed to the excessive noise for these planes.

Figure \ref{xcat_recon} shows results with the inclusion of random and scatter events in the simulation with varying intensity prior strengths.
We observe that in such a more realistic setting, the differences between the MLACF reconstructions and the ground-truth activity reconstruction become larger but are possibly within acceptable range.
Figure \ref{xcat_noise} shows MLACF results with varying levels of noise in the data.
Although there are variations in the MLACF reconstructions compared to the reference MLEM reconstruction, comparable activity distributions can be reconstructed with these noise levels.
Furthermore, the differences do not seem to decrease with increasing counts.

\subsection{Patient data}
Figure \ref{patient_snr} shows joint activity and attenuation estimation with a decreasing coincidence window of the data for attenuation estimation.
The difference images show improving agreement, most noticeable in the brain region, of the joint activity reconstructions with an increasing SNR of the measurements which the \quotes{residual} correction estimates are computed from.
The relative absolute difference of the MLACF reconstructions to the gold-standard was computed in a brain region (cyan activity contour) as 6.9, 5.8, 5.0, 4.8 and 3.4 $\%$ with the decreasing coincidence window, respectively.
Similarly, in the post-reconstruction attenuation volumes the relative differences to the reference CT attenuation reconstruction in a lung region (green attenuation contour) were 87.3, 83.9, 79.6, 75.6 and 57.9 $\%$ with the increasing SNR, respectively.
Similar results were obtained with a lower intensity prior strength, result not shown here.

Figure \ref{patient_recons} shows joint activity and attenuation estimation of MLACF for the patient data with varying strengths of the intensity prior.
Comparatively similar accuracies are observed in the patient data as was previously observed in the XCAT simulation results.
The differences are visually strongest in the bladder region where the scatter is quite often expected to be least accurate.
The difference images also reveal a moderately small amount of patient motion in the form of a head tilt between the CT and PET acquisitions.

\section{Discussion\label{discussion}}
An alternative to estimating the attenuation image in joint reconstructions from TOF-PET data is to estimate the attenuation factors which are then used to reconstruct the tracer activity distribution.
With the ever increasing axial extent of scanners and the decreasing crystal sizes, generating a histogram of the TOF events is becoming increasingly challenging.
For these reasons, this study works towards a solution for joint reconstructions using listmode emission data.
In this work we have deviated from the Poisson noise model and used a non-TOF model for estimating the AFs.
A disadvantage of this deviation is that we are no longer optimizing the same cost function for the estimated activity and attenuation pair, but importantly we benefit from the smaller memory demands, requiring only (mashed) non-TOF histograms of the emission data.
The differences of the two models are thought to be minimal for count levels where AF estimation becomes possible, provided that the emission data are histogrammed with \quotes{optimal} mashing to maximize the data SNR for joint reconstruction.

Furthermore, with the modified formulation of the listmode MLACF activity update, the estimated correction factors are directly used and a sensitivity image is generated from the measured events list as opposed to the collection of all possible LORs.
This helps avoiding the need for any pre-corrections to the emission data, mediates the computation-demanding step of computing the sensitivity image from all possible LORs, and only requires an additional non-TOF back/projection, making this listmode MLACF activity update more efficient than standard OSEM reconstruction from listmode data. 

In this work the attenuation factors were divided into 3 components; known tissue support, a residual factor for LORs intersecting media other than tissue and a scale factor. 
The scale factor was estimated post reconstruction and the known tissue support was computed from a preliminary non attenuation corrected reconstruction.
This allowed us to incorporate an intensity prior to favor factors without change during listmode MLACF, which helped limit any possible global fluctuations of the activity reconstruction within iterations.
In addition to this, the intensity prior also helps mitigating problems due to data-bins with limited information for joint reconstruction.
We found this not to be critical in simulations where the data is consistent, but more so needed in cases of possible inconsistencies within the data and to balance variations in the signal to noise ratio of the data which the AFs are computed from.

Since the problem of joint activity and attenuation reconstruction from TOF-PET is only determined up to a scale \cite{defrise2012j}, we propose to histogram the event-by-event \quotes{residual} correction factors to compute a scale factor.
A scaling factor in the AFs corresponds to the addition of a position/support dependent term in the  attenuation image. 
Applying this correction during the reconstruction of the attenuation is delicate and prone to inaccuracy.
That is why we propose a technique to compute a scale factor applied to the AFs which preserves the expected tissue attenuation in a known expected tissue region.
We find this method for solving the joint reconstruction scale to converge rapidly.
Furthermore, we have utilized a smoothing prior during the reconstruction of the estimated AFs.
However, this reconstruction is only used to compute the scaling factor and has no other effects on the final activity image.

As the estimation of the AFs is only possible along LORs that have sufficient counts, the results presented here are with a transaxial and axial crystal mashing of 4, combining the data for every four neighboring crystals.
This mashing is only used for getting a measure of the non-TOF counts/scatters for each measured event, and does not degrade quality of the activity reconstructions.
Visual assessment of the reconstructed attenuation volumes do not show immediate signs of resolution loss, however, a more systematic study of the effect of different mash strategies will be done in the future.

Results of the 3D Monte Carlo simulations revealed that accurate listmode MLACF reconstructions, compared to the ground truth OSEM reconstructions, can be obtained from TOF emission data only.
Close to the edges of the axial field of view however larger inaccuracies were observed.
We also observed that uncertainties in the expected additive randoms/scatter contributions add to the disagreement between the reconstructions.
We found however that a single scaling factor was sufficient for the quantification of the reconstructed activity image, as opposed to a plane-dependent scaling reported in \cite{bal2017j}.
This could be partly explained by the improvements in the time of flight accuracies between the studies and partly due to the presence of more oblique lines-of-response, which cause a stronger coupling between all image values, but requires further investigations.
The quantitative comparison of the patient data reconstructions showed good agreements in the (high uptake) brain region, however higher deviations were observed in the segmented lung region. 
We believe this to be partially due to respiration as the relative absolute differences of the lung region in the XCAT experiment was much smaller in comparison (14.9\%).

\section{Conclusion}
In this work we provide a reconstruction framework for the joint estimation of activity and attenuation from listmode TOF-PET emission data. 
The listmode MLACF algorithm alternately updates the activity reconstruction and a \quotes{residual} attenuation factors term.
It avoids the huge memory and computation demands for computing the AFs and the correspondingly evolving sensitivity image by using a Gaussian noise model on the estimated AFs and by avoiding the need to compute the sensitivity image from all possible LORs, respectively.
A post reconstruction of the estimated AFs, enforcing expected tissue attenuation values, provides the missing scale factor for a quantitative reconstruction.
The 3D simulations results reveal that the uncertainties in the additive contribution of the emission data increase differences compared to the ground-truth.
Nevertheless, the joint reconstruction results of the patient data are comparable to the gold-standard reconstructions using CT AFs.

\section*{Acknowledgement}
The authors would like to thank C.~Wiers from the University of Pennsylvania for the FDG patient study and M.~Werner for help in processing the data.
All authors declare that they have no known conflicts of interest in terms of competing financial interests or personal relationships that could have an influence or are relevant to the work reported in this paper.

\section*{Appendix}\label{delayed}
We express the likelihood for delayed events as a function of crystal singles rates $s$ as:
\begin{eqnarray}
   L^d = \sum_{\alpha \beta} d_{\alpha \beta} \ln{s_\alpha s_\beta} - s_\alpha s_\beta
   \label{MLdelayed}
\end{eqnarray}
where, $d_{\alpha \beta}$ is the number of delayed events recorded along the LOR connecting crystals $\alpha$ and $\beta$.
Approximating (\ref{MLdelayed}) with the same methodology as in \cite{rezaei2020j} the singles rates is computed as:
\begin{eqnarray}
   s^{m+1}_\alpha = s^m_\alpha (1 + \frac{\sum_\beta d_{\alpha \beta} - s^m_\alpha s^m_\beta}{2 \sum_\beta s^m_\alpha s^m_\beta})
   \label{singles}
\end{eqnarray}
where the super-script $m$ denotes the sub-iteration number.

\printbibliography

@article{pantel2019j,
	title = {{PennPET} Explorer: Human Imaging on a Whole-body Imager},
	issn = {0161-5505, 2159-662X},
	url = {http://jnm.snmjournals.org/content/early/2019/11/15/jnumed.119.231845},
	doi = {10.2967/jnumed.119.231845},
	shorttitle = {{PennPET} Explorer},
	pages = {jnumed.119.231845},
	journaltitle = {Journal of Nuclear Medicine},
	shortjournal = {J Nucl Med},
	author = {Pantel, Austin R. and Viswanath, Varsha and Daube-Witherspoon, Margaret E. and Dubroff, Jacob G. and Muehllehner, Gerd and Parma, Michael J. and Pryma, Daniel A. and Schubert, Erin K. and Mankoff, David A. and Karp, Joel S.},
	urldate = {2019-12-02},
	date = {2019-09-27},
	langid = {english},
	pmid = {31562224},
}

@article{badawi2019j,
	title = {First Human Imaging Studies with the {EXPLORER} Total-Body {PET} Scanner*},
	volume = {60},
	issn = {0161-5505, 2159-662X},
	url = {http://jnm.snmjournals.org/lookup/doi/10.2967/jnumed.119.226498},
	doi = {10.2967/jnumed.119.226498},
	pages = {299--303},
	number = {3},
	journaltitle = {Journal of Nuclear Medicine},
	shortjournal = {J Nucl Med},
	author = {Badawi, Ramsey D. and Shi, Hongcheng and Hu, Pengcheng and Chen, Shuguang and Xu, Tianyi and Price, Patricia M. and Ding, Yu and Spencer, Benjamin A. and Nardo, Lorenzo and Liu, Weiping and Bao, Jun and Jones, Terry and Li, Hongdi and Cherry, Simon R.},
	urldate = {2019-11-21},
	date = {2019-03},
	langid = {english},
}

@inproceedings{rezaei2015c,
	title = {Joint activity and attenuation reconstruction of listmode {TOF}-{PET} data},
	doi = {10.1109/NSSMIC.2015.7582250},
	pages = {1--3},
	booktitle = {{IEEE} Nuclear Science Symposium and Medical Imaging Conference Record ({NSS}/{MIC})},
	author = {Rezaei, A. and Bickell, M. and Fulton, R. and Nuyts, J.},
	date = {2015-10},
}

@article{rezaei2014j,
	title = {{ML}-reconstruction for {TOF}-{PET} with simultaneous estimation of the attenuation factors.},
	volume = {33},
	issn = {1558-254X},
	doi = {10.1109/TMI.2014.2318175},
	pages = {1563--72},
	number = {7},
	journaltitle = {{IEEE} Trans. Med. Imaging},
	author = {Rezaei, Ahmadreza and Defrise, Michel and Nuyts, Johan},
	date = {2014-07},
}

@article{nuyts2002j,
	title = {A concave prior penalizing relative differences for maximum-a-posteriori reconstruction in emission tomography},
	volume = {49},
	issn = {0018-9499},
	doi = {10.1109/TNS.2002.998681},
	pages = {56--60},
	number = {1},
	journaltitle = {{IEEE} Trans. Nucl. Sci.},
	author = {Nuyts, Johan and Beque, D. and Dupont, Patrick and Mortelmans, Luc},
	date = {2002-02},
}

@article{rezaei2017j,
	title = {Plane-dependent {ML} scatter scaling: 3D extension of the 2D simulated single scatter ({SSS}) estimate},
	volume = {62},
	issn = {0031-9155},
	url = {http://stacks.iop.org/0031-9155/62/i=16/a=6515},
	doi = {10.1088/1361-6560/aa7a8c},
	shorttitle = {Plane-dependent {ML} scatter scaling},
	pages = {6515},
	number = {16},
	journaltitle = {Physics in Medicine \& Biology},
	shortjournal = {Phys. Med. Biol.},
	author = {Rezaei, Ahmadreza and Salvo, Koen and Vahle, Thomas and Panin, Vladimir and Casey, Michael and Boada, Fernando and {Michel Defrise} and Nuyts, Johan},
	date = {2017},
	langid = {english},
}

@article{mehranian2015j,
	title = {Joint Estimation of Activity and Attenuation in Whole-Body {TOF} {PET}/{MRI} Using Constrained Gaussian Mixture Models},
	volume = {34},
	issn = {0278-0062},
	url = {http://ieeexplore.ieee.org/lpdocs/epic03/wrapper.htm?arnumber=7055275},
	doi = {10.1109/TMI.2015.2409157},
	pages = {1808--1821},
	number = {9},
	journaltitle = {{IEEE} Trans. Med. Imaging},
	author = {Mehranian, Abolfazl and Zaidi, Habib},
	date = {2015-09},
}

@article{defrise2012j,
	title = {Time-of-flight {PET} data determine the attenuation sinogram up to a constant},
	volume = {57},
	issn = {0031-9155},
	url = {https://iopscience.iop.org/article/10.1088/0031-9155/57/4/885/meta},
	doi = {10.1088/0031-9155/57/4/885},
	pages = {885},
	number = {4},
	journaltitle = {Physics in Medicine \& Biology},
	shortjournal = {Phys. Med. Biol.},
	author = {Defrise, Michel and Rezaei, Ahmadreza and Nuyts, Johan},
	urldate = {2019-04-19},
	date = {2012-01-31},
	langid = {english},
}

@article{vanslambrouck2014j,
	title = {Reconstruction Scheme for Accelerated Maximum Likelihood Reconstruction: The Patchwork Structure},
	volume = {61},
	issn = {0018-9499},
	doi = {10.1109/TNS.2013.2287637},
	pages = {173--181},
	number = {1},
	journaltitle = {{IEEE} Trans. Nucl. Sci.},
	author = {Van Slambrouck, Katrien and Nuyts, Johan},
	date = {2014-02},
}

@article{nuyts1999j,
	title = {Simultaneous maximum a posteriori reconstruction of attenuation and activity distributions from emission sinograms.},
	volume = {18},
	issn = {0278-0062},
	doi = {10.1109/42.774167},
	pages = {393--403},
	number = {5},
	journaltitle = {{IEEE} Trans. Med. Imaging},
	author = {Nuyts, Johan and Dupont, Patrick and Stroobants, S and Benninck, R and Mortelmans, Luc and Suetens, Paul},
	date = {1999-05},
	pmid = {10416801},
}

@article{rezaei2012j,
	title = {Simultaneous Reconstruction of Activity and Attenuation in Time-of-Flight {PET}},
	volume = {31},
	issn = {0278-0062},
	doi = {10.1109/TMI.2012.2212719},
	pages = {2224--2233},
	number = {12},
	journaltitle = {{IEEE} Transactions on Medical Imaging},
	author = {Rezaei, Ahmadreza and Defrise, M. and Bal, G. and Michel, C. and Conti, M. and Watson, C. and Nuyts, J.},
	date = {2012-12},
}

@inproceedings{panin2012c,
	title = {Reconstruction of uniform sensitivity emission image with partially known axial attenuation information in {PET}-{CT} scanners},
	isbn = {978-1-4673-2030-6},
	doi = {10.1109/NSSMIC.2012.6551495},
	pages = {2166--2173},
	booktitle = {2012 {IEEE} Nucl. Sci. Symp. Med. Imaging Conf. Rec. (2012 {NSS}/{MIC})},
	publisher = {{IEEE}},
	author = {Panin, Vladimir Y. and Defrise, Michel and Nuyts, Johan and Rezaei, Ahmadreza and Casey, Michael E.},
	date = {2012-10},
}

@article{rezaei2020j,
	title = {Estimation of Crystal Timing Properties and Efficiencies for the Improvement of (Joint) Maximum-Likelihood Reconstructions in {TOF}-{PET}},
	volume = {39},
	issn = {1558-254X},
	doi = {10.1109/TMI.2019.2938028},
	pages = {952--963},
	number = {4},
	journaltitle = {{IEEE} Transactions on Medical Imaging},
	author = {Rezaei, Ahmadreza and Schramm, Georg and Van Laere, Koen and Nuyts, Johan},
	date = {2020-04},
}

@article{badawi1999j,
	title = {Randoms variance reduction in 3D {PET}},
	volume = {44},
	issn = {0031-9155, 1361-6560},
	url = {http://stacks.iop.org/0031-9155/44/i=4/a=010?key=crossref.63d251a9a2ab65642c65f344cba04582},
	doi = {10.1088/0031-9155/44/4/010},
	pages = {941--954},
	number = {4},
	journaltitle = {Physics in Medicine and Biology},
	shortjournal = {Phys. Med. Biol.},
	author = {Badawi, R D and Miller, M P and Bailey, D L and Marsden, P K},
	urldate = {2020-02-27},
	date = {1999-04-01},
	langid = {english},
}

@article{li2020j,
	title = {Practical joint reconstruction of activity and attenuation with autonomous scaling for time-of-flight {PET}},
	volume = {65},
	issn = {0031-9155},
	url = {https://doi.org/10.1088/1361-6560/ab8d75},
	doi = {10.1088/1361-6560/ab8d75},
	pages = {235037},
	number = {23},
	journaltitle = {Physics in Medicine \& Biology},
	shortjournal = {Phys. Med. Biol.},
	author = {Li, Yusheng and Matej, Samuel and Karp, Joel S.},
	urldate = {2021-10-05},
	date = {2020-12},
	langid = {english},
	note = {Publisher: {IOP} Publishing},
}

@article{karp2020j,
	title = {{PennPET} Explorer: Design and Preliminary Performance of a Whole-Body Imager},
	volume = {61},
	issn = {0161-5505},
	url = {https://www.ncbi.nlm.nih.gov/pmc/articles/PMC6954465/},
	doi = {10.2967/jnumed.119.229997},
	shorttitle = {{PennPET} Explorer},
	pages = {136--143},
	number = {1},
	journaltitle = {Journal of Nuclear Medicine},
	shortjournal = {J Nucl Med},
	author = {Karp, Joel S. and Viswanath, Varsha and Geagan, Michael J. and Muehllehner, Gerd and Pantel, Austin R. and Parma, Michael J. and Perkins, Amy E. and Schmall, Jeffrey P. and Werner, Matthew E. and Daube-Witherspoon, Margaret E.},
	urldate = {2022-05-05},
	date = {2020-01},
	pmid = {31227573},
	pmcid = {PMC6954465},
}

@inproceedings{rezaei2022c,
	title = {Attenuation corrected activity reconstruction for listmode {TOF}-{PET} data},
	eventtitle = {{IEEE} Nuclear Science Symposium and Medical Imaging Conference ({NSS}/{MIC})},
	author = {Rezaei, Ahmadreza and Surti, Suleman and Karp, Joel S. and Nuyts, Johan},
	date = {2022},
}

@article{bal2017j,
	title = {Evaluation of {MLACF} based calculated attenuation brain {PET} imaging for {FDG} patient studies},
	volume = {62},
	issn = {0031-9155},
	url = {https://dx.doi.org/10.1088/1361-6560/aa5e99},
	doi = {10.1088/1361-6560/aa5e99},
	pages = {2542},
	number = {7},
	journaltitle = {Physics in Medicine \& Biology},
	shortjournal = {Phys. Med. Biol.},
	author = {Bal, Harshali and Panin, Vladimir Y. and Platsch, Guenther and Defrise, Michel and Hayden, Charles and Hutton, Chloe and Serrano, Benjamin and Paulmier, Benoit and Casey, Michael E.},
	urldate = {2024-10-08},
	date = {2017-03},
	langid = {english},
	note = {Publisher: {IOP} Publishing},
}

@article{lee2021j,
	title = {A Review of Deep-Learning-Based Approaches for Attenuation Correction in Positron Emission Tomography},
	volume = {5},
	issn = {2469-7303},
	url = {https://ieeexplore.ieee.org/abstract/document/9143173},
	doi = {10.1109/TRPMS.2020.3009269},
	pages = {160--184},
	number = {2},
	journaltitle = {{IEEE} Transactions on Radiation and Plasma Medical Sciences},
	author = {Lee, Jae Sung},
	urldate = {2024-10-09},
	date = {2021-03},
	note = {Conference Name: {IEEE} Transactions on Radiation and Plasma Medical Sciences},
}

@article{beck2009j,
	title = {A Fast Iterative Shrinkage-Thresholding Algorithm for Linear Inverse Problems},
	volume = {2},
	url = {https://epubs.siam.org/doi/10.1137/080716542},
	doi = {10.1137/080716542},
	pages = {183--202},
	number = {1},
	journaltitle = {{SIAM} Journal on Imaging Sciences},
	shortjournal = {{SIAM} J. Imaging Sci.},
	author = {Beck, Amir and Teboulle, Marc},
	urldate = {2024-10-16},
	date = {2009-01},
	note = {Publisher: Society for Industrial and Applied Mathematics},
}

\end{document}